# Data Integrity Verification in Network Slicing using Oracles and Smart Contracts

João Paulo de Brito Gonçalves
Federal Institute of Espirito Santo (Ifes)
Cachoeiro de Itapemirim - ES, Brazil
jpaulo@ifes.edu.br

Gustavo Alochio, Rodolfo da Silva Villaça, Roberta Lima Gomes
Federal University of Espirito Santo (Ufes)
Vitoria - ES, Brazil
gustavo.alochio@edu.ufes.br, rodolfo.villaca@ufes.br, rgomes@inf.ufes.br

*Abstract*—The fifth-generation (5G) wireless networks are expected to provide various services compared to the 4G and previous generations of networks. The Quality of Service requirements can be quite different in terms of latency, bandwidth, reliability, and availability. 5G technology allows the fragmentation of the network into small pieces, known as network slices. This network slicing is done by specific tools and the configuration must be protected from attacks that may be performed by malicious users. Thus in this paper, a solution to protect and prevent these failures from happening is addressed. For this solution to be carried out, a study was conducted on the Blockchain technology, as well as the use of Oracles in order to implement an integrity verification system, a system capable of assuring 5G network slices' configuration integrity through a complete architecture involving Blockchain, Smart Contracts and Oracles.

## I. INTRODUCTION

5G is one of the latest wireless networking technologies, and it is set to change how society lives. While 4G and 3G were very similar, 5G comes with a differentiated infrastructure, high data transmission speed and low latency, greater availability and network capacity [1]. Through this technology, a revolution will occur in several areas and some of these improvements will expand the research horizon, resulting in more capacity in urban mobility, process automation, transport infrastructure, hospitals, home automation, tourism and even agriculture.

With the new arrival and growing demand of using 5G networks in many places, new challenges also arise. The future of 5G networks is classified into two types, namely Public Land Mobile Network (PLMNs) or Private and Non-Public Network (NPNs). The first type is an infrastructure controlled by mobile network operators, and in the second type we have private networks, which are created for particular reasons, for specific environments and domains. 5G Network Slicing (NS) is a concept in which we have several categories of services and requirements addressed on the same network, with this slice being generated by a virtual addressing of the physical infrastructure. Network slices are proposed to control different categories of services and contexts within the same private network [1].

However, as the same slice provider can provide and manage slices for different organizations which may have conflicting and competing interests, mechanisms are needed to ensure the integrity of the system's data. In this context, this paper presents an integration between the 5G-EmPOWER tool [2] and the Ethereum blockchain, through Smart Contracts and Oracles, to guarantee the integrity of the configuration in the network slices.

This paper is organized as follows: in Section 2, we present the background concepts. In Section 3 we present the related work, Section 4 presents the development of the system initially proposed, as well its architecture and execution flow. In Section 5 the tests performed are presented and in the last section we present the conclusions and future work.

## II. BACKGROUND

### A. Blockchain

New blockchain networks emerged from the initial proposal of Satoshi Nakamoto [3]. In his work he proposed the blockchain technology and the first cryptocurrency, called Bitcoin. The blockchain Ethereum [4], known to be part of the second blockchain generation was designed with a different characteristic, the use of smart contracts in its network. Smart contracts are programs whose clauses are executed through transactions on the blockchain. From these new functionalities, the blockchain began to attract several supporters and promoters of technology, implemented in various areas, such as finance, government, industry and hospitals.

In public blockchains, i.e., where access is not controlled by a central authority, validation of transactions and blocks is often based on a consensus protocol. The initial consensus protocol proposed by Nakamoto was Proof of Work (PoW). In PoW, a cryptic challenge is proposed in order to create a valid block, once solved, the block is propagated over the network. Only after a transaction has been validated (included in a valid block), it is actually performed, which might change the blockchain state. PoW was proposed to discourage malicious users from creating fraudulent transactions on the network, but there is criticism regarding performance loss caused by their use. In the past few years, other validation strategies has been proposed, as Proof of Stake [5] and Proof of Authority [6] among others, based more in currency stake and reputation respectively and less in computation power.

### B. Ethereum and Smart Contracts

Ethereum [4] is a decentralized computing platform, which runs simultaneously on thousands of machines worldwide, which means that there is one single owner. So we can say that it is a decentralized platform with one or more nodes, and this number may depend on the purpose of the network. The Ethereum network can be of three types: Main Network, Test Network and Private Network.

Testnet networks are networks where generally applications run before been deployed in the Mainnet because they not require real monetary expenses. Private networks are those in which the nodes are not connected to any other Mainnet or Testnet participant. These networks can be used both for testing and for systems in production. There are also networks that can be maintained locally, these are only handled by the developer to speed up the testing of their applications, an example of a tool for providing local networks is Ganache [7].

As Mainnet and Testnet are public and non-permissioned networks, any user can participate and everyone can be nodes that participate in the consensus. Identification between the parties participating in a network transaction is usually done anonymously. The platform implements by default the consensus algorithm PoW.

In Ethereum there are two types of accounts: externally owned accounts (Externaly Owned Accounts - EOA) and contract accounts. EOAs are user accounts, while contract accounts are controlled by the code of a smart contract and executed by the EVM. The contracts, as they are accounts, also have addresses that identify them on the network, which can be used in transactions, as well as user accounts. However, contract accounts do not have a public or private key.

### C. Oracles

In the blockchain environment, oracles are systems that provide information from the real world to the blockchain. As the utilization of contracts evolved, some needs emerged, such as communicating with the world outside the network. As a result, smart contracts do not only deal with operations involving cryptocurrencies and internal blockchain data, but outside information may be involved in their system, which may include weather, stock prices, results and a range of options. Through the oracles a door on the blockchain opens to the outside world.

It was previously said that the blockchain was immutable, that is, transactions already carried out could never be changed or undone, but with the emergence of the concept of Oracles communication with the off-chain data became possible [8]. Oracles are decentralized services with one goal, to forward data from resources outside the blockchain network to the inside. The Ethereum documentation provides some references as Witnet [9], Provable [10], Paralink [10] and Dos.Network [11], but the most popular is the Chainlink [8] oracle network.

*1) Chainlink:* Chainlink [8] is a network of decentralized oracles that provide external data for smart contracts on the blockchain. The tokens LINK are the digital asset used to pay for services on the network.

A Chainlink network is composed of a collection of Chainlink nodes with registered Jobs specifications, which can perform Jobs execution coordinated by on-chain Oracle contracts, which use LINK Tokens as an incentive for Chainlink node operators to serve customers through Chainlink customer smart contracts. So, so far there are a few instances that need to be defined before getting an overview of how Chainlink works: the oracle smart contract, the client smart contract and the Node.

An oracle contract has functionality to allow the architecture that involves off-chain and on-chain to work. The first part controls which Chainlink clients also have invitations to fulfill work requests, another role would be to be a gateway of requested client nodes to the Chainlink network.

Chainlink nodes are responsible for executing the requested jobs. There may be a variety of Chainlink nodes connected to the blockchain network, which operate independently. They generally communicate with the blockchain node they are connected to, listening for work request events and sending results back via transactions.

The main consumer of Chainlink Oracles functionality is naturally the client smart contract, because it is not allowed to interact with the outside world. This has to be done through a mechanism where the need for the external data is conveyed by transaction events and an external party listening to these events is notified to the request and collects the requested data.

However, a single centralized oracle is a central point of failure. If the oracle is faulty or compromised, it would not be possible to know if the data is correct. Therefore, it is worthless if a smart contract uses oracles, but these provide unreliable data.

In this way, more than one oracle contract can be active in a blockchain network. A Chainlink node operator can choose which oracle contracts to register with to be allowed to handle work requests and a Chainlink client contract can choose which oracle to send the job request to. In Figure 1 it is possible to visualize the entire Chainlink structure.

### D. IPFS

The InterPlanetary File System (IPFS) [12] was created to allow the creation of a resilient and decentralized web, avoiding data centralization, as a request for data would not be directed specifically to a server, but to the decentralized network. Depending on the implementation, IPFS can have many advantages when compared to HTTP, such as censorship resistance, data integrity, lower operating costs, better performance and security. Some limitations are related to low incentives to participate in the network and its limited adoption that makes it difficult for files to be permanently available. If certain data is only hosted by a handful of nodes and everyone goes offline, it will be inaccessible.

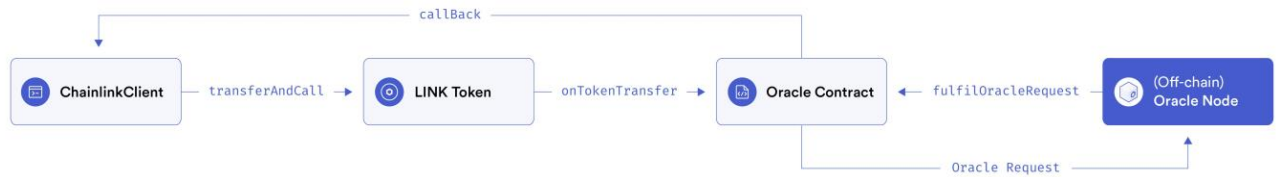

Fig. 1. Chainlink Architecture showing the Contracts, LINK Token, Node and the interactions between them

Like the blockchain, IPFS comes with the proposal to decentralize this data management, through nodes that contain fractions of the data inserted in the network, that is, a file when inserted in the network is partitioned and distributed by the nodes from the Web. Unlike the HTTP protocol, IPFS does not fetch data from just one source, but from several nodes, so it gains speed in this transfer. It requires less storage due to wide distribution, but also has great security due to the use of hash, being the files immutable, a user can check if the hash of the contents matches easily. The system guarantees decentralization with the use of nodes, avoiding unavailability. This type of hash is named as Content Identifier (CID). For content to remain active on the network, it needs to be pinned by its own node or by some third-party pinning service, otherwise the file may be deleted from the network as the system implements a garbage collector that removes files that are not being pinned.

### E. 5G-EmPOWER

A network slice consists of cross-domain components from separate domains in the same or different administrations. The concept of network slicing can facilitate multiple logical and independent networks on top of a shared physical infrastructure platform.

5G-EmPOWER [13] is a software defined networking tool (SDN) for 5G networks that provides slicing techniques. Its flexible architecture provides an open ecosystem where new 5G services can be tested under realistic conditions. 5G-EmPOWER uses specific techniques and hardware to virtualize WiFi access points and offer different quality of service (QoS) for each of the connected users. QoS is managed by dynamics, creating prioritization queues and directing data packets to the list that matches the required performance.

5G-EmPOWER is responsible for configuring WiFi access points, known by the acronym WTP (Wireless Termination Points). 5G-EmPOWER WTPs are based on a programmable hypervisor. The hypervisor controls the lifecycle of slices and is responsible for creating, monitoring, and managing them. Each Access Point (AP) can support a variable number of slices. On the other hand, each slice can handle multiple WiFi clients. A set of slices is called tenant. Among the parameters of a slice one crucial is the Quantum, which is the airtime (in ms) assigned to each slice, that represents the slice priority.

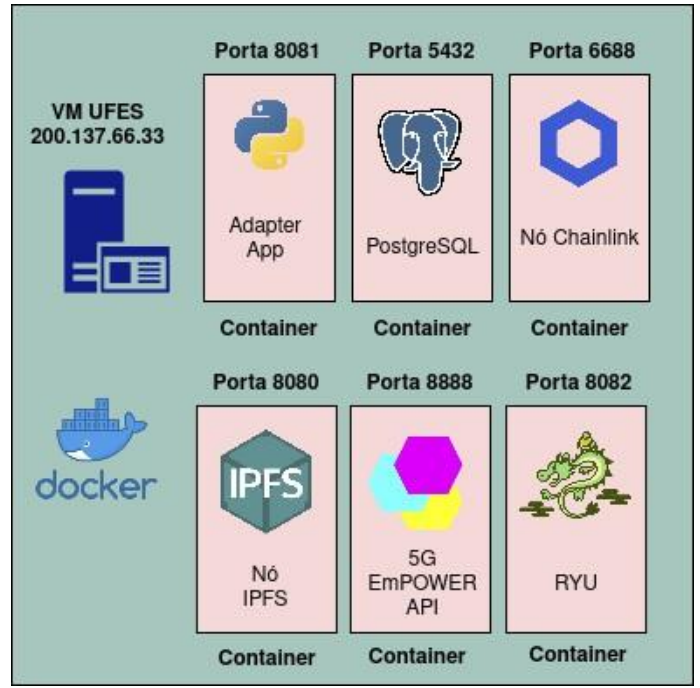

Fig. 2. Containers in the Virtual Machine showing the services and applications used in the implementation

### III. RELATED WORK

All the related work presented in this section share the same problem: the need to access data external to the blockchain that would be inaccessible without the use of some kind of oracle.

In [14] a framework to compare and characterize existing blockchain oracles mechanisms from industry is presented. The approach for reliability modelling and architecture analysis of blockchain oracle systems uses Fault Tree Analysis. Goncalves et. al. [15] propose an architecture to verify the levels of QoS in the services provided by a provider through an oracle developed with the Streamr framework [16] that receives and handles data streams in real time.

Buzachis et al. [17] and Azaria et al. [18] proposed solutions using smart contracts and blockchain to manage Electronic Health Records (EHR) and patient identities in e-health environments. A Proof of Concept was designed using the Ganache tool [7] that simulates a blockchain to simplify application deployment and tests, but without deployment in a real environment. MeDShare [19] and My Health My Data [20] are healthcare data systems that share data via cloud computing and blockchain technology.

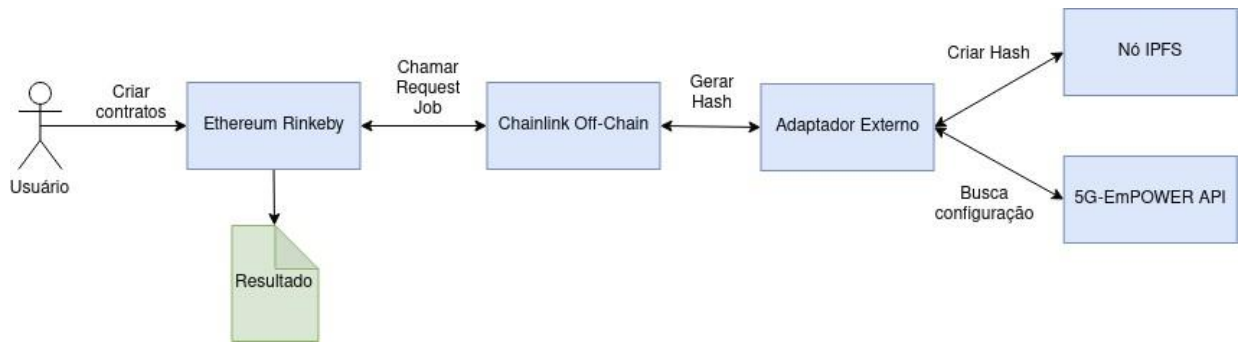

Fig. 3. Request Job Architecture showing the blockchain, Chainlink, IPFS and 5G-EmPOWER and the interactions between them

Pascale et al. [21] proposed a smart contract to automate Small-Cell-as-a-Service (SCaaS) agreements between the small-cell owners and network operators but without real evaluations or deployments. In Backman et al. [22] a Blockchain Network Slice Broker is proposed to reduce the service creation time for dynamically slice acquisition and for verifiable charging and billing in service level agreements, but no implementation or performance analysis is provided. Zanzi et al. [23] proposed NSBchain that uses the blockchain technology to address the new business models needs beyond traditional network slicing agreements. Afraz et al. [24] also proposed a 5G Network Slice Brokering. The idea is to use a distributed process to replace the conventional centralized approach to slice brokering, where a single authority does not control the entire conduct of the market.

## IV. DESIGN AND IMPLEMENTATION

To achieve the desired data decentralization and integrity, smart contracts technology was used in Ethereum blockchain with the addition of the use of Oracles. Through the Oracle it is possible to insert data external to the network and guarantee its immutability, as decentralized Oracles are reliable sources. It is still possible to guarantee access control to the contract's functionalities, thus guaranteeing security at the blockchain level.

With the data inserted in the decentralized network, it is possible to consult the contract and verify the veracity of the information: the integrity of the slices' data can be checked regarding the last authorized configuration change or if it was corrupted by an unauthorized modification.

To visualize the architecture of the system, it is necessary to understand how the tools are available for use and how they communicate with each other. For this, a virtual machine was used where all the necessary tools are placed in containers Docker, we can see this arrangement in Figure 2, where all the tools and services seen in this diagram are accessible through the Internet.

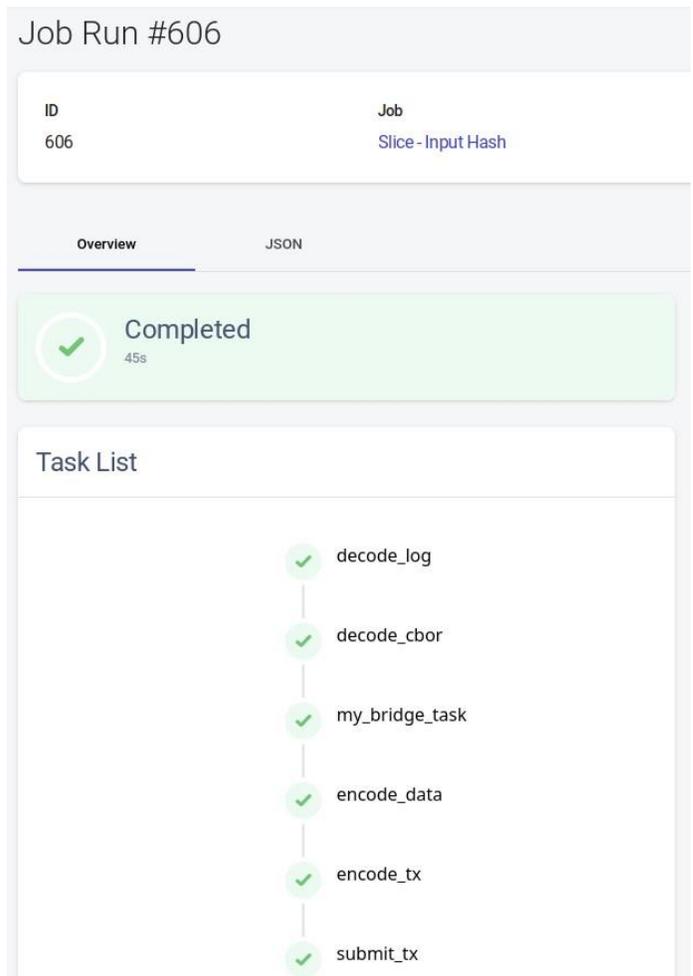

Fig. 4. Request Job Execution pipeline

The general idea that will be adopted in the proposed solution uses two large contexts, one from the Ethereum network and the other from Chainlink Oracle. In Ethereum, the smart contracts are the way in which the user will interact with the blockchain, and the hash information will be stored, information that will be automatically added by a call of the oracle. Initially, the smart contract does not have any data,

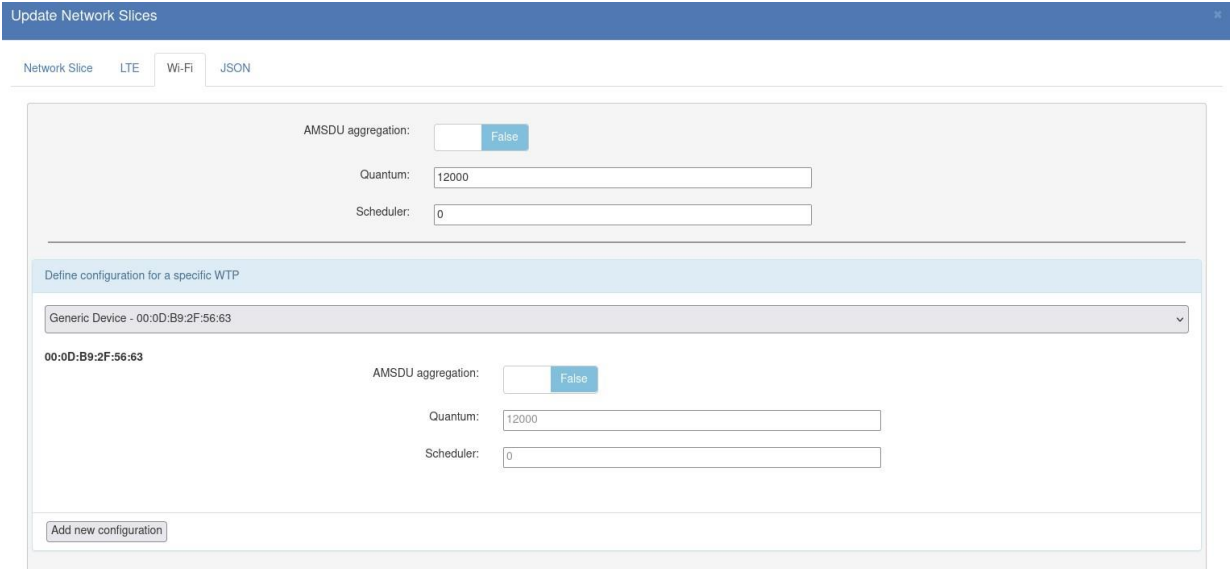

Fig. 5. Screen capture of the slice with identifier 0x00 and its quantum

```
[2022-03-24 22:59:16,271] DEBUG in app: Body: b'{"data":{"hashIpfs":"\\"QmRuqSaDTmvWQWhaY3RK5X8oxSJJpvzECZtk35gJfTzN6
e\\"\\n","path":"/api/v1/tenants/f7257cce-d05e-4f43-a0a6-f19236948f2f/slices/0x00"},"meta":{}}'
SUCCESS: Rota: /api/v1/tenants/f7257cce-d05e-4f43-a0a6-f19236948f2f/slices/0x00 verificada!
192.168.160.1 - - [24/Mar/2022 22:59:16] "POST /validate HTTP/1.1" 200 -
```

Fig. 6. Integrity Check Log showing that file has not been corrupted

```
[2022-03-24 23:00:36,182] DEBUG in app: Body: b'{"data":{"hashIpfs":"\\"QmRuqSaDTmvWQWhaY3RK5X8oxSJJpvzECZtk35gJfTzN6
e\\"\\n","path":"/api/v1/tenants/f7257cce-d05e-4f43-a0a6-f19236948f2f/slices/0x00"},"meta":{}}'
ERROR: Rota: /api/v1/tenants/f7257cce-d05e-4f43-a0a6-f19236948f2f/slices/0x00 corrompida!
192.168.160.1 - - [24/Mar/2022 23:00:36] "POST /validate HTTP/1.1" 200 -
```

Fig. 7. Integrity Check showing file corrupted after Quantum modification

only the path from where the information is accessed in the 5G-EmPOWER API. Using this API path, the oracle will get this information and send it to the IPFS system. The IPFS system will return the hash of the data stored in the API and the Oracle will take this hash back to the contract, thus signing this data inside the blockchain for further verification. The biggest motivation to use the IPFS system to save the collected information is due to the difficulty of handling large volumes of data in the blockchain, due to the high cost of storage and the time of validation of transactions within the Ethereum network.

For the oracle be able to perform these operations, an external adapter was developed inside a node Chainlink network node. The process begins with the client request, which requires a payment with an amount of LINK token. After the payment, a OracleRequest call is triggered for the contract that will receive the result of the transaction, which then triggers an event for the Chainlink Protocol, responsible for chaining the contracts creation such as, the Chainlink reputation contract that checks the reputation of the registered nodes, the contract for the delivery of the request and a last contract that takes all the data from the chosen oracles and validates and/or reconciles it for an accurate result.

## V. DEPLOYMENT AND EVALUATION

In order to validate the functionalities of the developed system, a use case test was carried out simulating an operation that can happen in a real environment. In this case, the following situation was evaluated: a user is an administrator of a 5G network slicing infrastructure through the 5G-EmPOWER platform and now wants to use the developed integrity system. This use is divided into two parts: one with the insertion of the hash of the valid configuration in the smart contract Validator, and another is the integrity verification of the data.

Thus, in the first step as seen in Figure 3, it creates a smart contract on the Ethereum Rinkeby network that has three parameters: path in the 5G-EmPOWER API, identifier of the Request Job in the Chainlink Node and the address of the Oracle contract, which must be created previously. With these three parameters inserted in the contract, the request can go beyond the blockchain barrier and make a call to Request Job. As a Chainlink Off-Chain operation, this operation will be listened to and performed by the Chainlink Node.

An interface for the Request Job operation is available in the dashboard of the Chainlink Node, and in figure 4 it is possible to visualize the list of tasks to be fulfilled by it. Among these tasks is my-bridge-task, being the task related to the IPFS External Adapter. The Request Job operation in the Chainlink Node propagates this call to the External Adapter, collecting the data in the API and delivering it to the IPFS system and thus requesting an IPFS CID.

The service Cron Job is responsible for verifying the integrity of the data comparing the data contained in the API to those contained in the IPFS hash. For this, the user creates a job of type Cron in the dashboard of the Chainlink Node and its execution pipeline. The Cron Job runs periodically (set for this test at 10 seconds).

The External Adapter Auditor is part of the Cron Job and its log is shown in Figure 6 with the scheduled data integrity verification and the result highlighted in red. After the first verification round, a change is made in the quantum parameter of the network slice configuration simulating a malicious privilege escalation, as we show in Figure 5. In the next scheduled data integrity verification step, the data will be checked, the modification is detected and reported as corrupted by the External Adapter Auditor, as shown in the log of Figure 7, also highlighted in red.

## VI. Conclusions and Future Work

This work had as its main goal the proposal of an architecture to verify the integrity of a 5G network slice configuration in a network slicing tool, and it was achieved through an implementation that integrated blockchain, smart contracts, oracles, 5G-EmPOWER and IPFS technologies.

For validation, the evaluated scenario was the use case where a 5G network slicing administrator creates a slice and needs to verify if is configuration data remain valid, with no change to it, easily. After the tests it was proved that it is possible to use the proposed system to verify the integrity of the configuration, validating our proposal.

As future work, some topics are suggested to improve this work, such as:

- An interface that allows the user to manage the contracts.
- Use a private blockchain network to implement the network.
- Evaluate the cost of transactions and propose other possibilities, alternatives, and feasibility.

## Acknowledgements

This research was financed in part by the following Brazilian research agencies: FAPES, FAPESP/MCTIC/CGI.br (#2020/05182-3) and IFES.